\documentstyle[12pt]{article} 

\newcommand{\beq}{\begin{equation}}
\newcommand{\eeq}{\end{equation}}
\newcommand{\dsl}{\raise.15ex\hbox{/}\kern-.57em\partial}
\newcommand{\dslash}{\,\raise.15ex\hbox{/}\mkern-13.5mu D}
\newcommand{\aslash}{\,\raise.15ex\hbox{/}\mkern-13.5mu A}
\newcommand{\pslash}{\raise.15ex\hbox{/}\kern-.57em p}

\begin{document}

\begin{center}
{\Large\bf  
Quantization of Chern-Simons Coefficient } \\[5mm]
Han-Ying Guo and Wan-Yun Zhao\\
Institute of Theoretical Physics, Academia Sinica\\
P.O.Box 2735, Beijing 100080, China
\end{center}

\vspace*{0.6cm}

\begin{center}
\begin{center} \bf Abstract \end{center}
\begin{minipage}{13cm}
The relation between the Dirac quantization condition of magnetic charge and 
the quantization of the Chern-Simons coefficient is obtained. It implies   
that in a (2+1)-dimensional QED with the Chern-Simons topological mass term 
and the existence of a magnetic monopole with magnetic charge $g$, the Chern-  
-Simons coefficient must be also quantized, just as in the non-Abelian case.\\[3mm]
PACS numbers: 11.10.Kk, 11.15.-q, 12.20.Ds
\end{minipage}
\end{center}

\vspace*{1cm}
             
  Recently, the quantization of Chern-Simons coefficient has attracted 
considerable attention [1]. The action with an added Chern-Simons topological
mass term to the usual Yang-Mills gauge theory in a (2+1)-dimensional space- 
-time remains invariant under small gauge transformations, but, to ensure   
invariance of the exponentiated action under large gauge transformations, 
the coefficient of Chern-Simons topological mass term has to be quantized [2].\par 
  In the Abelian case, however, in general, the Chern-Simons coefficient is
not quantized in the absence of a topological charge, but, the Chern-Simons 
coefficient must be also quantized in the presence of a topological charge,
for example, a magnetic pole, just as in the non-Abelian case.\par  
  In the letter we will demonstrate it. By the two-loop radiative corrections
perturbatively for the fermionic current vector in QED with a Chern-Simons
topological mass term in a (2+1)-dimensional space-time, we look for the
relation between the Dirac quantization condition of a magnetic charge and
the quantization of the Chern-Simons coefficient, thus the Dirac quantization
condition of a magnetic charge leads to the quantization of the Chern-Simons
coefficient. 

Let us consider the following Lagrangian in the (2+1)-dimensional QED with 
a Chern-Simons mass term,  
\beq
{\cal L}=-\frac{1}{4}F_{\mu \nu}^2-\frac{m}{2}\epsilon^{\mu\nu\lambda}A_{\mu}\partial_{\nu}{A_{\lambda}}
-\frac{\lambda}{2}(\partial_{\mu}A^{\mu})^2+\bar{\Psi}[\gamma_{\mu}(i\partial^{\mu}-e A^{\mu})-m_f]\Psi
\eeq
where $F_{\mu \nu}=\partial_{\mu}A_{\nu}-\partial_{\nu}A_{\mu};$ m, $\lambda$ and $m_f$ are the photon topological mass, a                   
parameter of the gauge-fixing term and the fermion mass respectively. The 
space-time is Minkowski with signature $(+, -, -)$. The $\epsilon^{\mu\nu\lambda}$ is a three-dimensional 
antisymmetric tensor. Our purpose is to find out the relation between the  
quantization of the Chern-Simons coefficient and the Dirac quantization 
condition of a magnetic charge in the presence of a magnetic pole in the 
Abelian case, therefore we can start with the electron current vector $J_{\mu}(x)$ 
in an external electromagnetic field $A_{\mu}(x)$ with the Chern-Simons topological  
mass term in 2+1 dimensions.
  
The ground state current vector of fermion fields in a (2+1)-dimensional  
space-time is
\beq
\langle0\mid J_{\mu}(x)\mid0\rangle=\langle0\mid\frac{1}{2}[\bar{\Psi}(x),\gamma_{\mu}\Psi(x)]_{-}\mid0\rangle                                 
=-Tr[\gamma_{\mu}G(x,x')]|_{x^{\prime} \rightarrow x}
\eeq 
where G(x,x') is the fermion propagator in interaction with an external
electromagnetic field $A_{\mu}(x)$ with the Chern-Simons mass term and it 
satisfies the following equation
\beq
(\dslash -m_f)G(x,x')=\delta^4(x-x')
\eeq
with
\beq
\dslash =i\dsl-e \aslash =\gamma_{\mu}[i\partial_{\mu}-eA_{\mu}(x)].
\eeq
  
What are considered are the effects which are brought by the Chern-Simons  
mass term, so the photon propagator $D_{\mu\nu}$ is taken into the following
form which is produced by the pure Chern-Simons term, i.e., the second term
of the Lagrangian in eq.~(1),
\beq
D_{\mu\nu}=\frac{\epsilon_{\mu\nu\lambda}k^{\lambda}}{mk^2}
\eeq
where it is in momentum space and we adopt the Landau gauge, $\lambda=0$, in eq.(1)  
in order to avoid the infrared divergences. In the (2+1)-dimensional
perturbative QED, the pure Chern-Simons effects brought by the fermion vector
current in eq.~(2) are in the two-loop corrections in the lowest order, so we 
have to consider the two-loop Feynman diagrams for the ground state current
vector of the fermion fields, $\langle0\mid J_{\mu}\mid0\rangle$, in a 
(2+1)-dimensional space-time in eq.~(2). These two-loop Feynman diagrams in 
momentum space are the diagrams (a), (b), (c) and (d) in Fig.~1.  

  The contribution of Fig.1(a) to the ground state current vector of the
fermion fields, $\langle0\mid J_{\mu}\mid0\rangle$, has the power of $e^3$, but the contributions of 
Fig.~1(b), (c), (d) to $\langle0\mid J_{\mu}\mid0\rangle$ have the power of $e^4$. After calculating the 
contribution of Fig.1(a) to the fermion current, we have discovered that it
vanishes. Therefore, in the perturbative QED of 2+1 dimensions, the lowest 
order of the nonvanishing contributions with the effect of Chern-Simons term
to the ground state current vector of the fermion fields is in O($e^4$), and
the corresponding Feynman diagrams are the three diagrams (b), (c), and (d)
in Fig.~1.

We now consider the contributions of the three diagrams Fig.~1(b), (c) and
(d). let us first take a look at the subdiagram in Fig.~1(b), the electron 
self-energy. To its contribution in momentum space
\beq
\Sigma(p)=\frac{e^2}{4\pi m}\{(p^2-m_f \pslash) \int_0^1\alpha^\frac{1}{2}[(1-\alpha)p^2+m_f^2]^{-\frac{1}{2}} d\alpha + 2m_f\} 
\eeq
where the Landau gauge has been adopted and the photon propagator has been
taken as the form in eq.~(5) due to the pure Chern-Simons theory, and the 
dimensional regularization has been also used. We compute the contribution
$\Pi_{\mu \nu}^{(b)}(k)A_{\nu}(k)$ of Fig.1(b) in terms of the electron self-energy $\Sigma(p)$ of eq.(6). Because
we concern with the pure Chern-Simons effects in $\Pi_{\mu \nu}^{(b)}(k)$, for simplicity, the electron mass
$m_f$ in $\Pi_{\mu \nu}(k)$ can be omitted, i.e., assuming the gauge boson mass $m \gg m_f$.
We have got for $\Pi_{\mu \nu}^{(b)}(k)$ after a tedious calculation
\beq
\Pi_{\mu \nu}^{(b)}(k)=\frac{e^4}{48\pi^2 m}[-\frac{2}{\varepsilon}-ln\frac{k^2}{\mu^2}-\gamma+O(\varepsilon)]\epsilon_{\mu \nu \lambda}k^{\lambda}
\eeq
where $\varepsilon=3-d$, the dimensional regularization has been used and the dimension
$d\rightarrow 3$, and $\gamma$ is Euler constant and $\mu$ is an arbitrary mass scale.\par
  The two diagrams (b) and (c) in Fig.~1 obviously give equal contributions.
We now turn to the diagram (d) in Fig.~1. Similarly, we proceed to the cumbrous
computation of the contribution $\Pi_{\mu \nu}^{(d)}(k)A_{\nu}(k)$ of 
Fig.~1(d) 
to the ground state vector current of the electron fields. We have also omitted
the electron mass $m_f$ in $\Pi_{\mu \nu}^{(d)}(k)$. We compute the contribution $\Pi_{\mu \nu}^{(d)}(k)$ in
terms of the Ward-Takahashi identities between the electron self-energy and the
vertex function which is in the subdiagram of Fig.~1(d). After a tedious computation
for $\Pi_{\mu \nu}^{(d)}(k)$ we have obtained 
\beq
\Pi_{\mu \nu}^{(d)}(k)=\frac{e^4}{12\pi^2 m}[-\frac{2}{\varepsilon}-ln\frac{k^2}{\mu^2}-\gamma+O(\varepsilon)]\epsilon_{\mu \nu \lambda}k^{\lambda}   
\eeq
where $3-d=\varepsilon$, $(k^2)^\frac{\varepsilon}{2}=1+\frac{\varepsilon}{2}lnk^2+ \cdot \cdot \cdot $ and  $\Gamma(\frac{3-d}{2})=-\frac{2}{\varepsilon}-\gamma+ \cdot \cdot \cdot $. Ultimately,
adding the contributions of the three diagrams in Fig.~1 to the fermion vector
current $J_{\mu}$, we have obtained the total contributions of Fig.~1(b), (c), (d) to
the electron vector current $J_{\mu}$ in momentum space therefore
\beq
J_{\mu}(k)=\frac{e^4}{8\pi^2 m}[\frac{2}{\varepsilon}+ln\frac{k^2}{\mu^2}+\gamma+O(\varepsilon)]\epsilon_{\mu \lambda \nu}k^{\lambda}A^{\nu}(k)
\eeq
\par  
  We now transform from a momentum space into a configuration space to consider
eq.~(9). In configuration space a magnetic field
\beq
\vec B=\vec\partial\times\vec A(x)=\epsilon_{ij}\partial_{i}A_{j}(x)
\eeq
where $\epsilon_{ij}=\epsilon_{0ij}, i,j=1,2$.\par
  The magnetic flux through the surface ( i.e., the two space dimensions) is
\beq
\oint\vec B\cdot d\vec\sigma=\int\vec B\cdot d\vec x=g
\eeq
by definition of the magnetic charge $g$ contained inside the sphere. There is
a magnetic monopole with a magnetic charge $g$.\par
  The contributions of the self-energy counterterms and the vertex counterterms
corresponding to Fig.~1(b), (c) and (d) can be also readily computed. After  
having considered the contributions of these couterterms and renormalized for
the quantities in eq.~(9), we finally get the result in configuration space
\beq
\int J_{0}(x)d\vec x=e_R=\frac{e_{R}^4}{8\pi^{2}m_R}\int\vec B\cdot d\vec x=\frac{e_{R}^{4}}{8\pi^{2}m_R} g
\eeq
where the zeroth dividual quantity $J_{0}(x)$ of $J_{\mu}(x)$ is the electric charge density,
and $e_R$ and $m_R$ are a renormalized electron charge and a renormalized Chern-Simons
topological mass respectively. This yields 
\beq
\frac{1}{2\pi}(e g)_R=n,
\eeq
\beq
4\pi(\frac{m}{e^2})_R=n,
\eeq
where $ n=0, 1, 2, 3, \cdot \cdot \cdot $.
Equation~(13) is called the Dirac quantization 
condition of magnetic charge. 
Equation~(14)
implies the quantization of the coefficient of Chern-Simons term. 
Thus it can
be seen that we have established the 
relation between the Dirac quantization
condition of magnetic charge and the quantization of the Chern-Simons
coefficient in QED of 2+1 space-time dimensions with the Chern-Simons 
topological mass term in the existence of any magnetic monopole with magnetic
charge $g$. It is easy to see from eq.~(12) that the Dirac quantization 
condition
of magnetic charge leads to the quantization of the Chern-Simons coefficient
or conversely, the requirement of the quantization of the Chern-Simons
coefficient in order to keep the nontrivial gauge invariance leads to the 
Dirac quantization condition of magnetic charge. Therefore, this demonstrate
that the Chern-Simons coefficient must be also quantized in the Abelian case
in the existence of a topological charge.

In this letter, in the (2+1)-dimensional QED with a Chern-Simons term the  
magnetic monopole is not put by hand, but is innate in the theory, so the 
quantization of the Chern-Simons coefficient is not also put by hand, but is 
also innate in the theory. The work of the quantization of the Chern-Simons
coefficient of the (2+1)-dimensional non-Abelian gauge theory at zero
temperature and finite temperature is in progress.

\begin{center}
\setlength{\unitlength}{1mm}
\begin{picture}(130,120)(20,60)
\put(0,0){\includegraphics{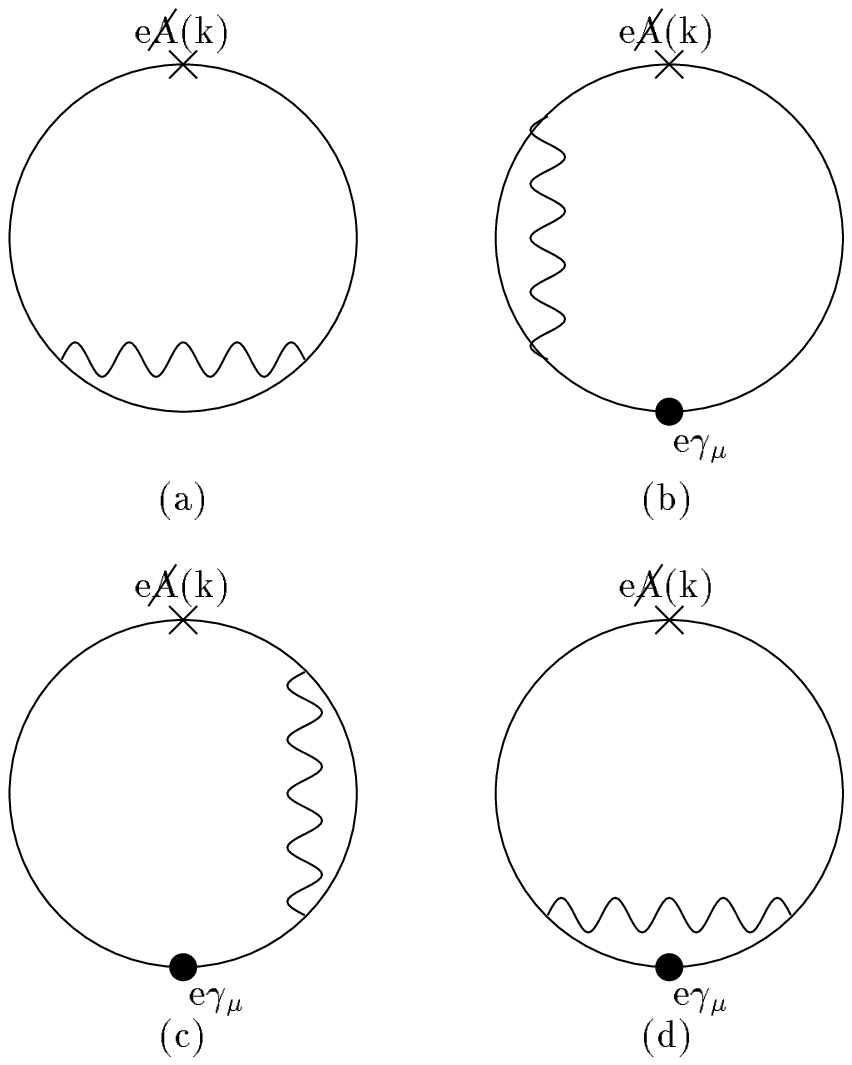}}
\end{picture}
\begin{minipage}{130mm}
Fig.~1. The two-loop Feynman diagrams for $\langle0\mid J_{\mu}\mid0\rangle$. 
The solid line stands for a fermion propagator, the wavy line stands for a 
gauge propagator in the pure Chern-Simons theory, $D_{\mu\nu}(k)$ in eq.~(5),
the cross for $e \aslash$ and the black dot for $e\gamma_{\mu}$.
\end{minipage}
\end{center}

\end{document}